**Failure of the Maxwell relation for the quantification of caloric effects in ferroic materials**

R. Niemann[a,b*], O. Heczko[c], L. Schultz[a,b] and S. Fähler[a,b]

[a]IFW Dresden, P.O. Box 270116, 01171 Dresden, Germany

[b]Department of Physics, Institute for Solid State Physics, Dresden University of Technology, 01062 Dresden, Germany

[c]Institute of Physics, Academy of Science of the Czech Republic, Na Slovance 2, 182 02 Prague, Czech Republic

* Corresponding authors email: r.niemann@ifw-dresden.de

Giant caloric effects were reported in elasto-, electro- and magnetocaloric materials near phase transformations. Commonly, their entropy change is indirectly evaluated by a Maxwell relation. We report the fundamental failure of this approach. We analyze exemplarily the Ni-Mn-Ga magnetic shape memory alloy. An applied field results in magnetically induced reorientation of martensitic variants, which form during the phase transformation. This results in a spurious magnetocaloric effect, which only disappears when repeating the measurement a second time. This failure is universal as the vector character of the applied field is not considered in the common scalar evaluation of a Maxwell relation.

Solid state refrigeration based on magnetocaloric[1], elastocaloric[2], or electrocaloric[3] effects is considered as a promising route for energy efficient cooling. Giant effects are reported in vicinity of phase transformations, resulting in an abrupt change of extensive properties like spontaneous (magnetic or electric) polarization or strain. These symmetry-reducing transformations in addition lead to the formation of anisotropic entities like magnetic or ferroelectric domains and martensitic variants, which are switchable by external (magnetic, electric or stress) fields. The entropy changes as a measure of the cooling efficiency is commonly evaluated indirectly by means of Maxwell relations.[1-3]

Here we demonstrate the failure of this approach. As an example we analyze an apparent magnetocaloric effect of magnetic shape memory alloys, which are multiferroic alloys exhibiting both, a ferromagnetic, and ferroelastic phase transformation. We will show that this failure is universal since the vector character of the applied field is ignored in the scalar evaluation of a Maxwell relation.

A Maxwell relation of Gibb's Free Energy G allows obtaining the entropy change *ΔS* as a function of temperature from isothermal measurements of strain, polarization or magnetization curves at different temperatures T:

$$\Delta S = \int_{\Delta Y} \left(\frac{\partial X}{\partial T}\right)_Y dY \qquad (1)$$

where X is the strain, polarization or magnetization M and Y the stress-, electric or magnetic field H, respectively. In particular for a magnetocaloric material the entropy change is given by an integrated and discretized form of the Maxwell relation to Gibb's Free Energy[4,5,6]

$$\Delta S\left(\frac{T_{i+1}-T_i}{2}, H\right) = \frac{\mu_0}{(T_{i+1}-T_i)} \int_0^H [M(T_{i+1}, H')-M(T_i, H')] dH' \qquad (2)$$

when subsequent magnetization curves M(H) are measured in discrete temperature steps $\Delta T = T_{i+1}-T_i$. In the following we will call this indirect measurement the 'standard procedure'.

Rare-earth and MnAs-based magnetocaloric materials usually exhibit entropy changes of several -10 Jkg$^{-1}$K$^{-1}$ near room temperature.[7] As recently analyzed by Caron et al.[8] under some circumstances the determined magnetic entropy changes exceed the theoretical magnetic limit given by $\Delta S_M^{max} = R\ln(2J+1)$, ref.[9], where R is the universal gas constant and J the total angular momentum of the Mn ion. For instance, in Mn$_{0.99}$Cu$_{0.01}$As a maximum entropy change of -178 Jkg$^{-1}$K$^{-1}$ was determined using eq. (2) while the magnetic limit $\Delta S_M^{max}$ in this system[10] is -103 Jkg$^{-1}$K$^{-1}$. Caron et al.[8] identified the hysteresis of a first order transformation as origin of this discrepancy between two phases exhibiting the required difference in magnetization in combination with a small ΔT. They suggested circumventing this problem by adjusting the measurement sequence. Cooling the sample well below the phase transformation temperature before starting each M(H) measurement allows compensating the hysteretic behavior. This is called the 'loop procedure'. For the particular example of $Mn_{0.99}Cu_{0.01}As$ this method gave a maximum entropy change of 78 Jkg$^{-1}$K$^{-1}$, a value well below $\Delta S_M^{max}$.

As an example that the proposed loop procedure is incomplete we examine a sample of the Ni-Mn-Ga magnetic shape memory alloys. In these alloys substantial magnetocaloric effects up to -20 Jkg$^{-1}$K$^{-1}$ in a magnetic field change from 0 to 1.6 T are obtained when composition is selected that Curie temperature and martensitic transformation temperature coincide.[11,12] However, this value was determined using the indirect method. In contrast to that a direct temperature change of 1.2 K was observed in $Ni_{55}Mn_{20}Ga_{25}$ under the adiabatic application of 1.6 T at 320 K. This corresponds to an entropy change of only $\Delta S \cong -\frac{c_p \Delta T}{T}$ = -1.54 Jkg$^{-1}$K$^{-1}$. Additionally in small fields an inverse magnetocaloric effect of 4 Jkg$^{-1}$K$^{-1}$ was reported in the vicinity of the martensitic transformation which was ascribed to the strong temperature dependency of the magnetocrystalline anisotropy.[13,14]. Planes et al. already identified an extrinsic contribution to the magnetocaloric effect which originates from the martensitic variants in Ni-Mn-Ga[15]. We will clarify how this extrinsic contribution yields an apparent magnetocaloric effect when the material is analyzed by isothermal magnetization measurements using eq. (2).

In addition to magnetocaloric effects, magnetic shape memory alloys exhibit a magnetically induced reorientation (MIR) within the martensitic state.[16] During MIR the martensitic variants are aligned with their easy axis along the external magnetic field. This effect can be used for magnetic actuation with strains up to 10%[17] and requires high magnetocrystalline anisotropy[18]. Since applicable magnetocaloric devices could only operate economically using commercial permanent magnets as field source, we limit our investigations to maximum field changes of 1 T.

For the present experiments we used a $Ni_{50}Mn_{28}Ga_{22}$ single crystal with an almost cubic shape with faces approximately cut along {100} planes. The martensitic crystal structure is 5M, where the short c-axis of the almost tetragonal crystal structure is the easy magnetization axis. Differential scanning calorimetry (DSC, Fig. 1) at 4 Kmin$^{-1}$ was used to determine the martensitic transformation temperatures as $A_S = 322$ K, $A_F = 333$ K, $M_S = 318$ K, and $M_F = 311$ K. These measurements reveal a total entropy change of the martensitic transformation of -9.6 Jkg$^{-1}$K$^{-1}$. Except for a change in slope at the Curie transition $T_C = 369$ K no further features are observed in the DSC curve.

Since both, MIR and the magnetocaloric effect, are probed by M(H) measurements, we first summarize the characteristic stages of MIR in Fig. 2. Measurement had been performed in a Quantum Design PPMS with vibrating sample magnetometer option at 300 K. As a starting point for the experiments, we compressed the single crystal mechanically, which results in the formation of a single variant state with the short c-axis aligned along the stress. We mounted the sample such that the easy magnetization axis is perpendicular to the external field. Hence in low fields a hard axis magnetization loop is observed, where the magnetization vector coherently rotates towards the field direction. When the switching field $H_{SW} = 0.36$ T is reached, a crystallographic variant with the easy axis in field direction starts to grow on the expense of the initial variant. This variant has the easy axis along the field direction, hence magnetization increases strongly. When the magnetic field is removed there is no need for further changes of orientation, so we call this reorientation an irreversible process.[19] For the following temperature dependent measurements one has to consider that $H_{SW}$ increases with decreasing temperature.[20]

To determine the entropy change, we first used the standard procedure: The sample was cooled to 250 K and consecutive $M(H)$ loops up to 1 T in steps of 5 K were measured (Fig. 3a, only a selection of loops is shown for clarity). At 250 K, only the linear increase of a hard axis loop is observed since we apply the field perpendicular to the easy magnetization axis. The following measurements, repeated in increasing temperature, change gradually and exhibit a two-part behavior with an initial steep part up to 0.2 T and a rather flat part up to the maximum field. This is the consequence of the temperature dependence of the MIR effect[20]: At low temperatures, the field required to reorient a part of the sample exceed the anisotropy field, so no MIR occurs. With increasing temperature the switching field decreases and a part of the sample can reorient. MIR occurs close to the anisotropy field, since here the driving energy for MIR is maximal. In contrast to the first measurement at one temperature (Fig. 2), no abrupt jump in magnetization is observed, because the sample is always almost saturated when the reorientation occurs. We attribute the observed spread of switching fields over a broad temperature range to microstructural inhomogeneities exhibiting different pinning efficiency. The transformation is not complete until 300 K. This reorientation process is irreversible since no external force will re-align the variants to the original state. Further on, when the temperature is increased towards $A_S$ and $T_C$, the common decrease of spontaneous magnetization is observed.

These magnetization curves had been evaluated according eq. (2) and the resulting entropy change is plotted in Fig. 4. Close to $A_S$ the common magnetocaloric effect of -1.1 Jkg$^{-1}$K$^{-1}$ is observed. Additionally in the temperature range from 267 K to 302 K, where MIR occurs, an apparent inverse magnetocaloric effect with a maximum entropy change of $-0.8$ Jkg$^{-1}$K$^{-1}$ at 287.5 K is obtained. In the temperature region below 300 K the DSC measurement exhibit no features, which excludes any phase transformation. This implies that this remarkable inverse magnetocaloric effect is spurious.

Next we repeated these measurements following the loop procedure. In order to avoid an influence of thermal hysteresis, we cooled the sample to 250 K before each M(H) loop. As with the first series, we defined the initial variant distribution by mechanical compression at the beginning of this series and mounted the sample with the easy axis perpendicular to the external field. In this measurement cycle all measurements below 290 K show an almost linear hard-axis behavior until 290 K, where a small jump in magnetization occurs in a field of 0.94 T (Fig. 3b). This jump occurs due to an almost complete MIR since all following measurements reveal easy-axis behavior. We attribute difference of MIR compared to the standard procedure to a slightly different sample mounting (e. g. some restoring force due to thermal shrinkage of the fixating Teflon tape during undercooling). For a true magnetocaloric effect, this however should not play a role. The evaluation using eq. 2 shows that the non-physical inverse peak (Fig. 4) is now confined to a single data point as expected from MIR behavior while the common magnetocaloric effect in this sample is shifted down by 5 K and has a slightly higher absolute value of -1.2 Jkg$^{-1}$K$^{-1}$.

The effect of the MIR results in an apparent inverse magnetocaloric effect that is even bigger than the real common magnetocaloric peak, which demonstrates that the approach of Caron et al.[8] is incomplete. However, these irreversible effects can easily be avoided when repeating all magnetization measurements again without restoring the initial variant configuration. These measurements are plotted in Fig. 3c. Since during the first cycle the easy axis was already aligned along the external field, now only a steep increase of magnetization is observed. Extracting the entropy change according to eq. 2 (Fig. 4) shows that the spurious peak at 290 K vanishes; only the common low magnetocaloric effect of -1.2 Jkg$^{-1}$K$^{-1}$ in vicinity of the martensitic transformation remains.

To conclude, we demonstrated that the vector character of the applied field can result in a re-arrangement of the anisotropic entities formed at a diffusionless transformation. This can result in hysteresis and irreversible processes which are not considered in the scalar application of a Maxwell relation. This indirect method can result in spurious non-physical magnetocaloric effects. The present experiments give an intuitive description of the problem: MIR does nothing else but rotating the (almost cubic) Ni-Mn-Ga single crystal by 90°. Since the entropy cannot depend on the orientation of the system or the direction the examiner looks on it, the magnetocaloric effect derived indirectly from magnetization measurements is obviously an artifact of the measurement procedure. In order to avoid irreversible processes and obtain relevant physical data one should repeat these measurements a second time and use the second measurements only.

These spurious effects are a fundamental problem, since most diffusionless phase transformations result in anisotropic, switchable entities: martensitic variants, magnetic or ferroelectric domains. It is not limited to the present case of well-trained single crystals, for instance MIR is also observed in polycrystals[21] and antiferromagnets[22]. Finally we will illustrate in a thought experiment, that spurious 'colossal' effects are expected also for materials with a single ferroic transition of second order. For this we consider hard magnetic materials like Nd-Fe-B where a magnetic field only changes the magnetic domain configuration. In a demagnetized sample with random magnetic domain structure, the first, virgin magnetization curve $M(H > 0)$ at a temperature $T_0$ shows a gradual increase of magnetization until saturation is reached in a field of about 1 T. In a second magnetization curve at $T_0 + 1$ K an almost constant magnetization $M(H > 0) \cong M_S$ is expected due to the high remanence of a hard magnet (all magnetic domains are aligned along the field). Re-evaluating e.g. a magnetization measurement of ac-demagnetized Nd-Fe-B[23] using eq. 3 gives an apparent "colossal" entropy change of ≈ 280 Jkg$^{-1}$K$^{-1}$. This un-physical effect would vanish in a second measurement.

The authors acknowledge funding by DFG through SPP 1239 (Grant No. FA 453/7) and by ASCR through an International cooperation grant.

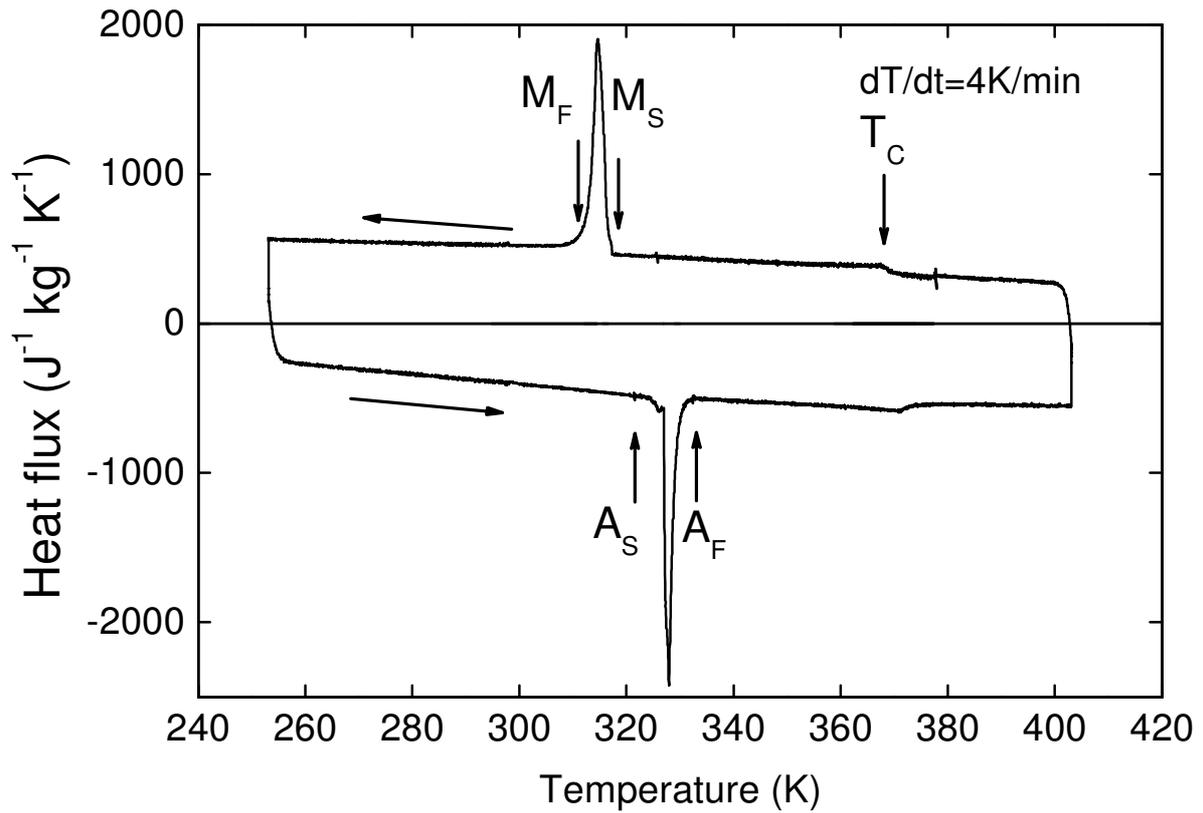

Fig. 1. The heat flux from the sample was measured as a function of temperature measured by differential scanning calorimetry. The peaks during cooling and heating are a consequence of the first-order martensitic and reverse martensitic transformation, respectively. The second-order Curie transition appears as a change in the slope in a small temperature interval. The hysteresis in $T_C$ is due to the large size of the sample.

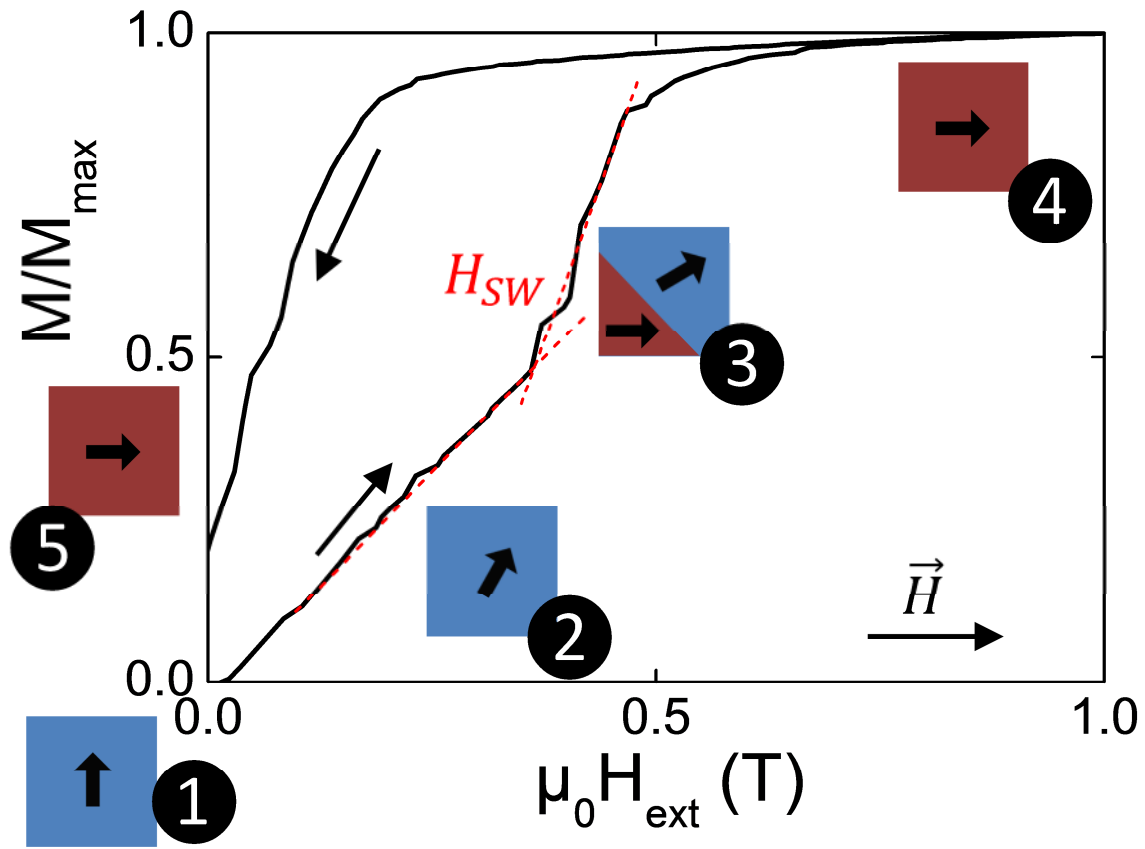

Fig. 2. Relation between magnetically induced reorientation (MIR) and measured magnetization curve in a $Ni_{50}Mn_{28}Ga_{22}$ single crystal measured at 300 K. The sketches describe the reorientation of the martensitic microstructure by twin boundary motion. ❶ Initially, a single-variant state (blue) was prepared by mechanical compression. ❷ The magnetic field was applied perpendicular to the easy axis which leads to a rotation of the magnetization. ❸ The jump in magnetization above the switching field $H_{SW}$ originates from the growth of a variant with the easy axis parallel to the applied field (brown). ❹ In high fields, the initial variant vanished and magnetisation saturates. When the field was removed, the sample remained in a single-variant state (brown). ❺ Due to the macroscopic symmetry of the cube-shaped crystal initial and end-stage of the experiment is equivalent to a rotation of the sample by 90°.

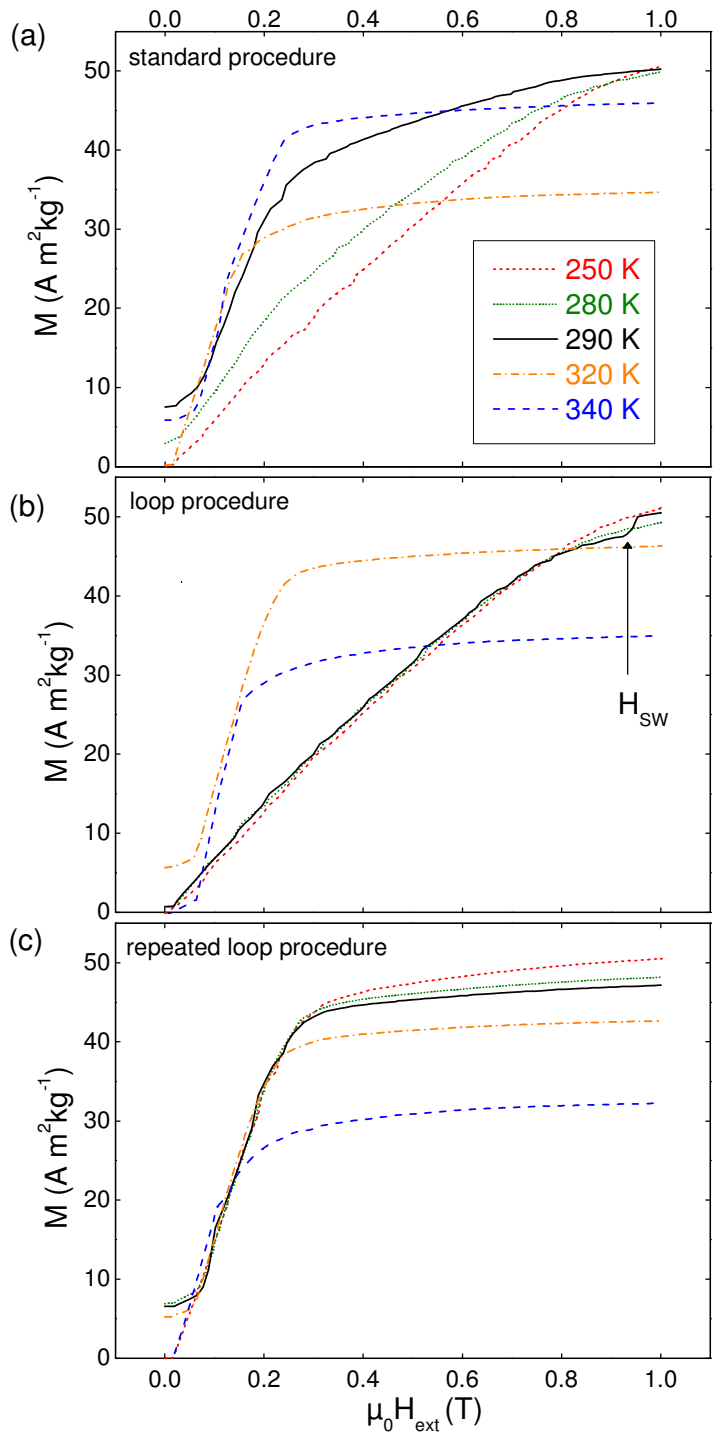

Fig. 3. Magnetization curves as a function of the increasing applied magnetic field measured in steps of 5 K from 250 K to 340 K (for clarity, only a selection is shown).
(a) Standard procedure: Prior to the first magnetization measurement, the sample was transformed mechanically into a single variant state. The magnetic field was applied perpendicular to the easy axis of this single variant. (b) Loop procedure: As with the first procedure the field was applied perpendicular to the easy axis of a single variant. Additionally, the sample was cooled to 250 K prior to each temperature step. (c) Repeated loop procedure: the previous measurement was repeated without bringing the sample initially into a single variant state.

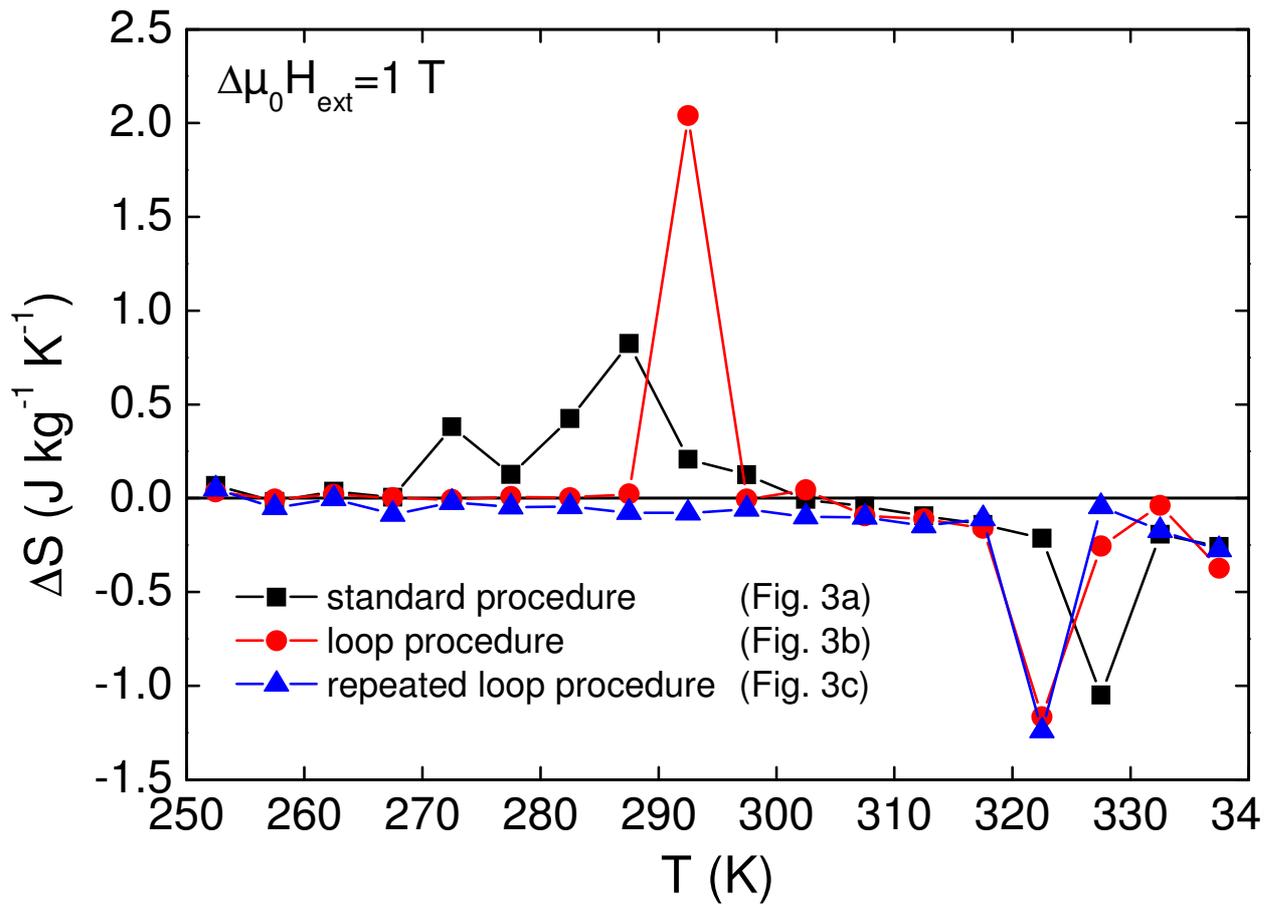

Fig. 4. Entropy change as a function of temperature in a change of the magnetic field from 0 to 1 T calculated from the three different sequences of magnetization measurements (Fig. 3a-c) using equation (2). While the minimum around 325 K reflects the change in magnetization during the (reverse) martensitic transformation, the peak around 285 K is a spurious inverse magnetocaloric effect originating from the irreversible magnetically induced reorientation processes.